# Observation of Nanometer-Scale Rolling Motion Mediated by Commensurate Contact

M. R. Falvo[1,4], J. Steele[1,4], A. Buldum[1,4], J. D. Schall [2,4], R. M. Taylor II [3,4],
J.P. Lu[1,4], D. W. Brenner [2,4], R. Superfine[1,4]

[1]*Dept. of Physics and Astronomy, The University of North Carolina, Chapel Hill, NC 27599-3255*
[2]*Dept. of Materials Science and Engineering, North Carolina State University, Raleigh, NC*
[3]*Dept. of Computer Science, The University of North Carolina, Chapel Hill, NC*
[4]*North Carolina Center for Nanoscale Materials, The University of North Carolina, Chapel Hill*

We report, through experimental observations and computer simulations, that atomic lattice interlocking can determine whether an object rolls or slides on a surface. We have quantitatively manipulated carbon nanotubes (CNTs) on a variety of substrates with an atomic force microscope (AFM) and observe rolling to occur only on graphite. We measure the forces when the CNT is in-registry with the graphite lattice, and observe rolling only in this lock-in state. Atomistic computer simulations identify the energy barriers for sliding and rolling, elucidate atomic-scale features of slip-roll motion, and explain the details of the lateral force data in terms of the intrinsic faceting of multiwall CNTs.

The interaction between two bodies in contact is ultimately determined by the interaction between atoms. Understanding how these interactions affect energy loss [1-4] and object motion is important for designing lubrication strategies and self-assembly processes, and will determine the forms of atomic-scale actuating devices[5]. Current microelectromechanical devices have features typically in the size scale of ten microns, and gears have been fabricated with teeth measured in the same size range. It is of great interest to understand the ultimate scale of actuating devices, and in what manner atomic interactions will play a determining role [6]. In the present work, we show through both experiment and simulation that the interlocking of the atomic lattices in the contact region of two bodies can determine whether the body slides or rolls. In essence, the atomic lattice can act like a gear mechanism.

The arrangement of the atoms in two interacting surfaces has been shown to play a critical role in the energy loss that occurs when one body slides over a second both in experiment [7-9] and simulation [10-12]. More generally, the motion of a free body can involve several degrees of freedom, including sliding, rotating in the plane,[13, 14] and rolling. As a model system for such studies, CNTs offer a regular geometry with atomically smooth surfaces that can remain relatively clean in ambient laboratory conditions. We have performed experiments that demonstrate that multiwall CNTs roll on graphite (HOPG) only when the atomic lattices of the CNT and the substrate are in registry. Our calculations quantitatively describe these observations and show how intrinsic deformations of CNTs may affect atomic scale mechanical devices.

The CNTs, prepared by the arc-discharge method [15], were sonicated in ethanol then dispersed and evaporated onto HOPG. AFM [16] manipulations, performed in ambient, employ an advanced operator interface called the nanoManipulator (nM). [17, 18] During each manipulation, the calibrated lateral force [19, 20] is monitored as a measure of the CNT substrate friction. As the AFM tip is pushed into contact with the CNT in a trajectory perpendicular to its axis, the CNT undergoes either in-plane rotation or rolling [21]. The CNTs move as rigid bodies, which is expected for CNTs of this size (10-50nm diameter, 500-2000nm length) considering their high stiffness and the low friction of the graphite substrate [22]. If the CNT starts in an out-of-registry state, it slides smoothly [21]. However, this motion is interrupted at discrete in-plane orientations where the CNT "locks" into a low energy state,[23, 24] indicated by a ten-fold increase in the force required to move the CNT (Fig. 1A). Subsequently, the CNT rolls with a constant in-plane orientation and characteristic stick-slip modulation in the lateral force [21] (other examples of stick-slip lateral force traces characteristic of rolling motion are shown in Fig. 4(a)). Most important, CNTs can be manipulated to reveal a set of in-registry orientations separated by 60+/-1 degrees (Fig. 1A inset). The lock-in angles and the associated increase in lateral force are reproducible for a given tube. This evidence shows that the CNT is registering with the hexagonal lattice of the the graphite substrate.

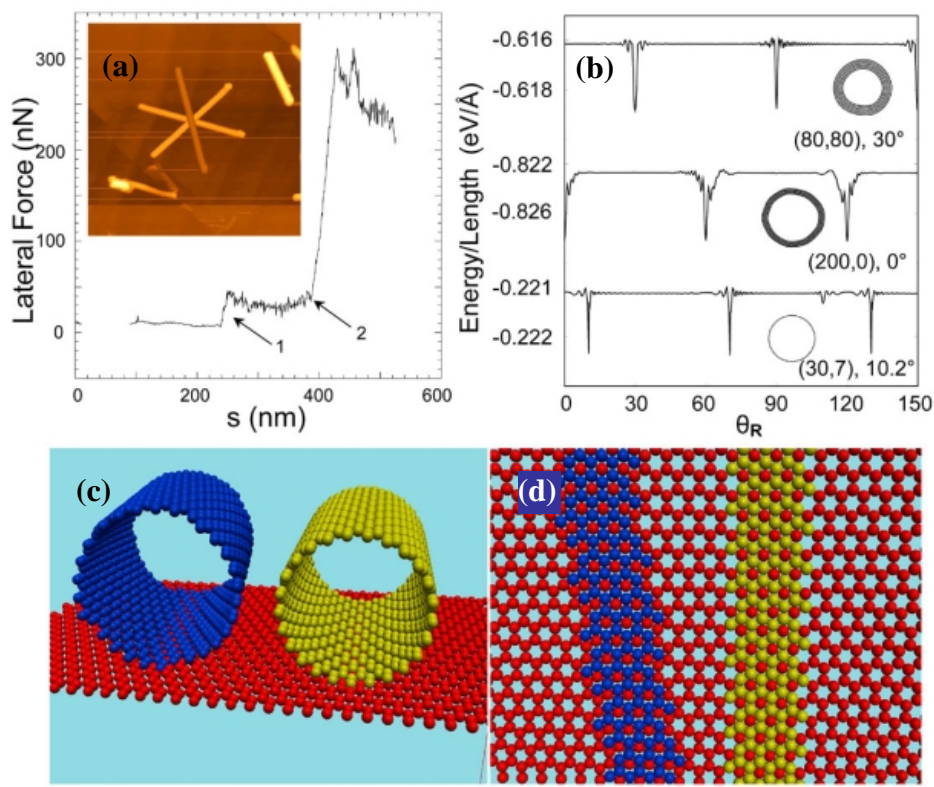

Figure 1. (A) Lateral force trace during an AFM manipulation showing the large change in force when a CNT goes from an out-of-registry orientation into an in-registry state. At point 1, the AFM tip contacts the CNT and begins sliding/rotating the tube. At point 2, the tube reaches the registered orientation, the lateral force increases almost 10 fold, and rolling motion begins. Three registered orientations are found for each CNT, 60 +/- 1 degree relative to each other. The inset shows three overlain AFM images of the locking orientations of a 20nm diameter, 960nm long CNT. (B) The interaction energy as a function of rotation angle between the CNT axis and the HOPG lattice for (80,80), (200,0) and (30,7) CNTs. Each CNT has unique minimum energy orientations repeating every 60°. The insets indicate the relaxed cross sectional shape of each CNT ignoring substrate interaction. Note that the (80,80) (6 walls, diameter = 10.9nm) and the (200,0) (5 walls, diameter-=15.7nm) CNTs have multiple walls and have non-circular cross sections as determined by MD calculations. The (30,7) (1 wall, diameter=2.67nm) CNT retains a circular cross section when relaxed. (C) Model of two nanotubes resting in commensurate contact on a HOPG surface (red). The blue tube is a (25,5) CNT and the yellow is a (25,0) CNT. (D) Shows the contact zone of the commensurate lattices. The tube axes of the two CNTs are 11 degrees relative to each other when in commensurate contact. This model of the two tubes is shown to stress the point that tubes of differing chiralities will have differing orientations of the tube axis when in commensurate contact.

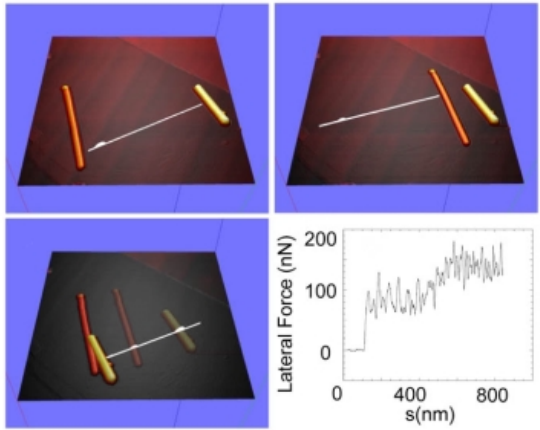

Figure 2. Manipulation of CNT on HOPG. The 3-fold locking axes for the tubes are off by 11 degrees relative to each other. (a) The white streak indicates the trajectory of the AFM tip during the manipulation, from lower left toward upper right. Both CNT are pushed from lower left to upper right (a-b). After (b), both CNTs are individually pushed back across the same area and as (c) indicates, the shorter CNT has been pushed into contact with the longer CNT such that both tubes translate further (the fainter CNTs show the original position and the bold CNTs show the final position). Figure (d) shows the lateral force during this manipulation.

To demonstrate that the atomic lattice of the CNT plays an equal role in the lock-in behavior, multiple CNTs have been manipulated at the same location on the graphite substrate (Fig. 2). In one of several similar manipulations, we have moved two CNTs (left CNT- 950nm long, 20nm diameter, right CNT - 500nm long, 34nm diameter) lying in the same immediate area on the graphite substrate. While each CNT shows the complete set of lock-in behaviors described above, the two CNTs have lock-in orientations that differ by 11 degrees. The sequence shows a series of manipulations in which the tubes are rolled individually across the same region in order to verify that the difference in their orientations is not due to an inhomogeneity in the graphite substrate. This difference in lock-in angles implicates the CNT lattice. If the registered orientations are due to atomic registry, the particular set of registered orientations is determined by the CNT chirality (the wrapping orientation of the outer graphene sheet of the CNT). Large multi-wall CNTs of different diameters are expected to have different chiralities [25] and should show different registered orientations as confirmed by our molecular statics calculations described below. Fig 1C and D show a model of two nanotubes of differing chiralities lying in commensurate contact with a relative angle between tube axes of 11 degrees.

Another manipulation emphasizes the robust gear-like motion of CNTs in the atomically registered state. We have manipulated two CNTs into a collision to observe the subsequent motion (Fig. 2(c)). The lateral force trace (Fig 2(d)) shows characteristic periodicity for the first tube before the collision, then an increase in the lateral force after the collision. Both CNT remain in their commensurate orientations after the collision.

We have performed atomistic simulations [26] to understand the hierarchy of CNT motions (rolling, in-plane rotation and sliding), and their dependence on CNT size, chirality and type (multiwall vs. singlewall). We first analyze the equilibrium positions and the energy barriers related to the motion by calculating the variation of total potential energy for a given rotational, translational or rolling position. We represent the interaction between a CNT and the graphite surface by an empirical potential of the Lennard-Jones type. This model has been used successfully to calculate the bulk properties of $C_{60}$ solids [27, 28]. The potential energy as a function of rotation angle (Fig. 1B) has deep minima for certain rotation at 60° intervals, which are directly related to the chiral angles of the CNTs. In all three cases, the deep minima correspond to atomic registry: orientations where the graphene sheet of the CNT and the graphene sheet of the substrate are in graphitic ABA stacking.

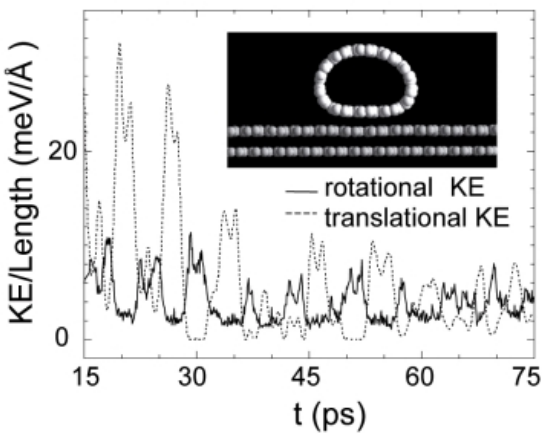

Figure 3. Molecular dynamics simulation of rolling (10,10) CNT in registry. After 15ps, rolling motion begins and continues until the CNT stops at 100 ps. This rolling motion consists of alternating sliding and spinning regimes, which is indicated by the alternating peaks of the rotational and translational kinetic energy in the figure. The inset shows the relaxed cross section of the (10,10) tube as it lies on the HOPG substrate.

Molecular dynamics simulations [29] were carried out to investigate the importance of lattice registry for the rolling behavior. In an initial simulation, a (10,10) CNT was placed in an in-registry starting position where the hexagonal lattice of the substrate matched that of the CNT. In a second simulation, the substrate was oriented 90 degrees to the CNT (out-of-registry). Note the significant deformation of the CNT (Fig. 3). A uniform force was applied to the CNTs for 0.5 ps to set them into motion. The simulation was then run until the CNTs stopped moving relative to the substrate.

Analysis of animated sequences from the simulations reveals that the out-of-registry CNT slides along the substrate and comes to a stop at ~ 30 ps without ever rolling. In contrast, the in-registry CNT initially slides, then begins a slide-spin motion, finally coming to a stop at 100ps. We focus on the alternate slide-spin motion that begins after approximately 15 ps of sliding. Fig. 3 shows the alternating rotational and translational kinetic energy after the CNT has begun this complex motion. The period of this alternation in distance (4.58 Å) corresponds to the graphite lattice repeat distance (4.26 Å) and the subperiod indicated by the double peak (1.54 Å) corresponds to the carbon-carbon bond length (1.42 Å). The CNT is "rolling over" to avoid direct atom-atom collisions (this is clearly observed in the animated sequence). The net effect of the spin-slide alternation is a form of rolling. This motion clearly has an energy loss rate per unit distance lower than the early time in-registry sliding (over several lattice spacings). The CNT loses over 80% of its energy during the first five lattice spacings as it slides, while the remaining 20% of energy is sufficient to have the tube roll for the last 5 lattice spacings.

The comparison between our simulations of perfect graphene cylinders and our experimental observations leaves several questions. For example, our molecular statics calculations on perfect graphene cylinders for a range of sizes (including the objects of our experiments), predict a *lower* force for the initiation of rolling than for other motion, whether in-registry or out-of-registry. We observe rolling only for the in-registry case. Furthermore, our observed experimental rolling forces are about ten times *larger* than in-plane rotation forces (Fig. 1) whereas our simulations predict the opposite ordering. The lateral force traces for a rolling CNT show a periodicity equal to the CNT circumference (Fig 4), with the trace shape reproducing in detail during each rolling period. If these characteristics arise from features on the CNT, are the features intrinsic to large multiwall CNTs or are our CNTs defective? Finally, why are the lateral force traces different for a given CNT rolling in the forward and backward directions? These features cannot arise from a perfect cylindrical tube.

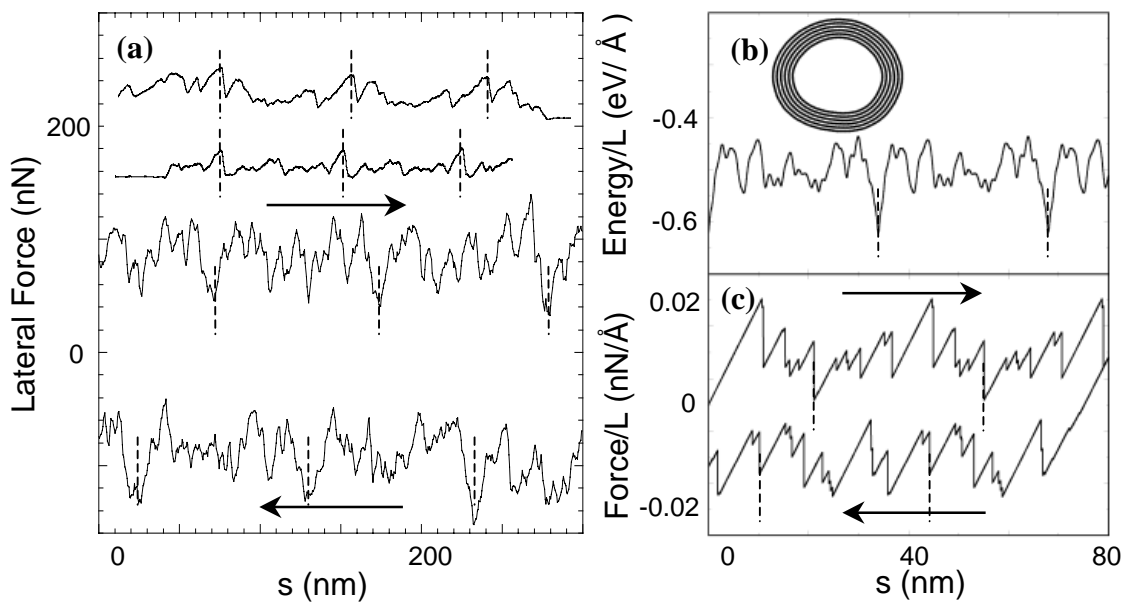

Figure 4. Rolling and Faceting. In (a) experimental lateral force traces are shown for three different rolling CNTs (from top to bottom the traces represent pushes of CNTs having diameters of 27, 29 and 35nm respectively). In each case, a period is apparent that is equal to the circumference of the CNT. The bottom two traces of (a) are for forward and reverse rolling of the 35nm CNT. Negative forces indicate the AFM cantilever is deflected in the opposite direction. The inset of (b) shows the structure of a multi-wall CNT as calculated with MD. The energy trace (b) shows the potential energy as a function of rolling distance for a CNT in atomic registry. The corrugation shows the energy cost of the lifting of a face up off the substrate and rolling onto the next face. The force trace in (c) shows a simulation of an AFM cantilever (modeled as a perfect spring of k = 50 N/m) pushing the CNT of (c) along the potential.

We performed molecular dynamics calculations (relaxation through simulated annealing) that show that a sufficiently large multiwall CNT is faceted. This is consistent with experimental evidence from in transmission electron microscopy,[30, 31] and neutron diffraction studies [32]. The CNT in the present calculation (Fig. 4B inset) is a multi-wall CNT having 6 layers (outer shell has chirality (80,80)). This faceting balances the lower potential energy due to the perfect graphitic stacking in the flattened faceted regions with the cost of high curvature between flat regions. We believe that this faceting plays a key role in our experimental observations. First, the energy cost for rolling now includes a component due to the adhesion of the facet face and the substrate, substantially larger than that for the perfect cylinder. This energy cost for rolling is lower than the energy cost of sliding only when the CNT is in registry. We have performed a simulation of the rolling of a faceted CNT with a spring force representative of an AFM cantilever (Fig.

4C). The stick-slip features show peak magnitudes in the lateral force per unit length of ~20 pN/Å, consistent with our experimental observations which range from 7 - 30 pN/Å. The rolling energy cost of for a perfectly cylindrical graphene tube of similar size is about 100 times smaller. Most important, the features of the lateral force in the simulation of rolling are periodic with the CNT circumference, and are clearly correlated with facets of the CNT. Furthermore, the reverse rolling force trace is different from that of forward rolling, consistent with the hysteretic instabilities expected for a spring/corrugated -potential system.

Our results have important implications for understanding nanoscale dynamics and the design of devices. For perfect graphene cylinders, our calculations show that atomic lattices can act as robust gear mechanisms. Contact area and CNT deformations play a critical role in the resulting dynamics. While for perfect cylinders rolling is always preferred, faceting or deformation can completely suppress rolling, even in the in-registry orientation. Our results indicate that the energy loss in the rolling of large diameter CNTs arises from the adhesion of faces, similar to a peeling process. Interesting questions remain in the energy loss mechanisms of rolling in perfect cylinders, friction at high velocities and the role of deformations in the dynamics of single wall CNTs and tube-tube junctions.

We thank the Otto Zhou for providing the CNT material, Sean Washburn for his important insights, and the whole nanoManipulator team for their invaluable work. This work was supported by the National Science Foundation (HPCC, ECS), the Office of Naval Research (MURI), and National Institutes of Health (NCRR).